\documentstyle[psfig,conf_iap]{article}

\begin{document}

\heading{Distribution of the Lyman-$\alpha$ Forest at $z = 0.0-1.7$}

\par\medskip\noindent

\author{Adam Dobrzycki$^{1}$, Jill Bechtold$^{2}$}

\address{Harvard-Smithsonian Center for Astrophysics, 60 Garden
Street, MS 70,\\
Cambridge, MA 02138, USA}

\address{Steward Observatory, University of Arizona, Tucson, AZ 85721,
USA}


\begin{abstract}
We present the analysis of a sample of the Ly-$\alpha$ forest spectra
of 152 quasars taken with the HST FOS. The Ly-$\alpha$ lines show
little evolution at $0<z<1.7$. We see a difference between the
evolution indices for weak and strong lines.
\end{abstract}


\section{Introduction}

We present the analysis of the distribution of Ly-$\alpha$ clouds at
$z \lsim 1.7$. Our line sample was drawn from the HST FOS spectra of
152 QSOs. In order to achieve self-consistency we reduced all the
spectra from scratch. The line sample consists of lines between the
Ly-$\beta$ and Ly-$\alpha$ emission lines. We excluded: metal systems
and regions adjacent to them, lines within 3000~km/s from the QSO
emission redshift, and lines with $W>1$~\AA.  The resulting sample has
595 lines.

The evolution of the Ly-$\alpha$ forest lines is usually described
with $d{\cal N}/dz = {\cal A}_0 (1+z)^\gamma$. For a non-evolving
population of clouds $\frac{1}{2} \leq \gamma \leq 1$, depending on
$q_0$. At $z>1.7$, $\gamma$ is significantly higher than 1, indicating
strong evolution (e.g.\ \cite{Bec1994} and references therein). The
results from the HST Key Project showed that for low redshifts
$\gamma$ is smaller, and in principle consistent with no evolution
(\cite{Bah1993}, \cite{Bah1996}; see \cite{Jan1997} for completed Key
Project results). We estimated $\gamma$ for various subsets of our
sample.

\section{Results}

See \cite{Dob1997} for detailed description of the sample and the
analysis. We present the major results in Tab.~1 and on Fig.~1.

We confirm that line density evolution at $z\lsim 1.7$ is much slower
than at higher redshifts and that it is consistent with the no
evolution case. It is possible that the number density of weak
lines increases with decreasing redshift. In any case, there is a
significant difference in the evolution of weak and strong lines.
While number densities of strong and weak lines are comparable at
$z\approx0$, the former outnumber the latter more than twice at
$z\approx1.5$. This can be an indication that the weak and strong
lines are members of separate populations of the absorbers.


\begin{figure}[t]
\centerline{\vbox{
\psfig{figure=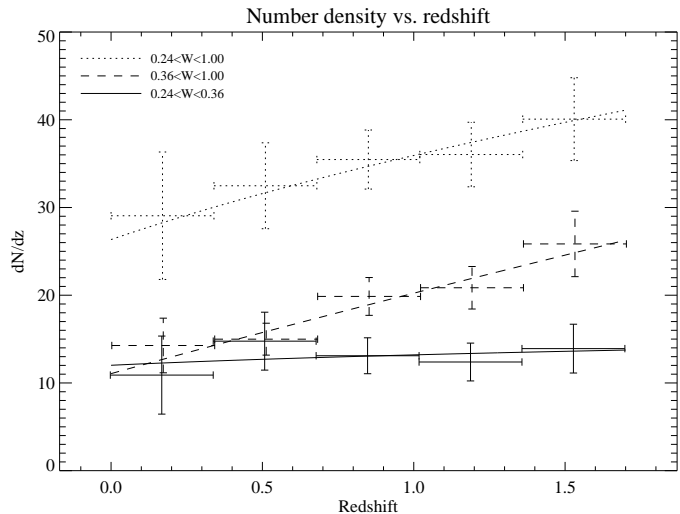,height=7.cm}
}}
\caption[]{Number density evolution for various subsamples. Dotted
lines: whole sample, dashed lines: absorbers with $W>0.36$~\AA, solid
lines: absorbers with $0.24<W<0.36$~\AA. Note distinctly different
slopes for weak and strong lines.}
\end{figure}


\begin{table}
\begin{center}
\scriptsize
\begin{tabular}{ccccc}
\hline\hline
\noalign{\vspace*{0.1ex}}
$W_{\rm min}$ [\AA] &
$W_{\rm max}$ [\AA] &
${\cal N}$ &
$\gamma$ &
${\cal A}_0$\\
\noalign{\vspace*{0.1ex}}
\hline
\noalign{\vspace*{0.1ex}}
$\cdots$ & 1.00  & 595 & $0.80\pm0.20$ & $\cdots$ \\
  0.24   & 1.00  & 339 & $0.45\pm0.28$ &   26.3   \\
  0.36   & 1.00  & 296 & $0.87\pm0.27$ &   11.0   \\
  0.24   & 0.36  & 125 & $0.14\pm0.45$ &   12.0   \\
\noalign{\vspace*{0.1ex}}
\hline\hline
\end{tabular}
\caption{Estimated evolution indices for various subsamples.}
\end{center}
\end{table}


\acknowledgements{We would like to thank D.~Dobrzycka, J.~Scott,
K.-V.~Tran, and B.~Wilden for their contributions to this project.
This research has made use of the NASA/IPAC Extragalactic Database
(NED) which is operated by the JPL, Caltech, under contract with the
NASA. This work is based on observations made with the NASA/ESA Hubble
Space Telescope, obtained from the data archive at the STScI, which is
operated by the AURA, Inc. under the NASA contract NAS~5-26555.
Support from STScI grant No.\ AR-05785.02-94A is acknowledged. AD is
supported by NASA Contract No.\ NAS8-39073. JB is supported by NSF
grant AST-9058510.}


\vfill

\end{document}